\documentclass{article}

\usepackage[
  paperheight=8.5in,
  paperwidth=5.5in,
  left=10mm,
  right=10mm,
  top=20mm,
  bottom=20mm]{geometry}
\usepackage[utf8]{inputenc}

\usepackage{graphicx}
\usepackage{wrapfig}
\usepackage[bottom]{footmisc}
\usepackage{listings}
\usepackage{enumitem}

\usepackage{wrapfig}
\usepackage{ragged2e}

\usepackage{array}
\usepackage[table]{xcolor}
\usepackage{multirow}
\usepackage{booktabs}
\usepackage{hhline}
\definecolor{palegreen}{rgb}{0.6,0.98,0.6}

\usepackage{amsmath}
\usepackage{amssymb}
\usepackage{multicol}
\usepackage{lipsum}
\usepackage{hyphenat}
\PassOptionsToPackage{hyphens}{url}
\usepackage{url}

\usepackage{rotating}

\usepackage[T1]{fontenc}
\usepackage{textcomp}
\usepackage{listings}
\usepackage{setspace}

\newcommand*{\email}[1]{\small{\texttt{#1}}}

\renewcommand{\footnoterule}{%
  \kern -3pt
  \hrule width \textwidth height 0.5pt
  \kern 2pt
}

\date{}

\usepackage{titlesec}
\titleformat*{\section}{\large\bfseries}
\titleformat*{\subsection}{\normalsize\bfseries}
\titleformat*{\subsubsection}{\normalsize\bfseries}

\usepackage{biblatex}

\title{Hands-on PDC in Undergraduate Computing Education\footnote{\protect Copyright \copyright 2025 by the Consortium for Computing Sciences in Colleges.
Permission to copy without fee all or part of this material is granted provided
that the copies are not made or distributed for direct commercial advantage,
the CCSC copyright notice and the title of the publication and its date appear,
and notice is given that copying is by permission of the Consortium for
Computing Sciences in Colleges.  To copy otherwise, or to republish, requires
a fee and/or specific permission.
}
}

\author{
Hala ElAarag and Anas Gamal Aly\\
Mathematics \& Computer Science\\
Stetson University\\
DeLand, FL 32723\\
\email{\{helaarag, agamal\}@stetson.edu}
}

\begin{document}
\maketitle

\begin{abstract}
Parallel and Distributed Computing (PDC) is a critical yet conceptually challenging area of the undergraduate computer science curriculum. While students often encounter these concepts in theory, few gain exposure to experience in real high-performance computing (HPC) environments. Research shows that when students are engaged in project-based learning they retain knowledge more effectively. They also develop a deeper understanding of concepts taught in the classroom. This paper presents a practical assignment in which students engage directly with the University of Florida’s HiPerGator supercomputer to implement and benchmark matrix multiplication using Python and C (via POSIX threads and OpenMP). Students navigate batch scheduling, core allocation, and performance tuning, experiences that are rarely accessible at the undergraduate level. We describe the assignment in detail and provide a three-year evaluation across multiple course offerings, highlighting how structured access to real HPC infrastructure can deepen student understanding of parallelism and multithreading.
\end{abstract}

\section{Introduction}

Undergraduate courses in Parallel and Distributed Computing (PDC) rarely grant students hands-on access to high-performance computing (HPC) resources. To bridge this gap, we developed an assignment in which students implement and benchmark multithreaded matrix-multiplication code on the University of Florida’s HiPerGator supercomputer~\cite{university2016hipergator,cook2011supercomputers}.  
Matrix multiplication is a $\mathcal{O}(n^{3})$ operation~\cite{stothers2010complexity} that underlies many advanced systems—most prominently the Transformer architecture that powers modern large-language models (LLMs)~\cite{vaswani2017attention}. Although recent research seeks to eliminate computationally-intensive matrix multiplication operations in deep learning~\cite{zhu2024scalable}, conventional LLM pipelines as of 2025 remain dominated by matrix multiplication~\cite{kachris2025survey}.

CS pedagogy studies show that hands-on, project-based activities increase engagement and long-term retention, especially for abstract topics such as concurrency and parallelism~\cite{lewandowski2005fostering,baldwin1996discovery}. When students tackle open-ended problems on real hardware—whether breadboards, Raspberry Pis, or HPC clusters — they report greater satisfaction and achieve deeper conceptual understanding~\cite{elaarag2022hands,elaarag2017deeper,pucher2011project}. Discovery learning is therefore potentially useful for PDC~\cite{adams2021teaching}.

Previous efforts have helped students visualize multithreaded behaviour~\cite{10.1145/2839509.2844557}. We extend this work by moving beyond simulation: students write POSIX-threaded and OpenMP implementations, craft batch scripts, submit jobs to HiPerGator, and analyse scalability across languages, thread counts, and cores. The assignment situates algorithmic theory within a real-life HPC workflow, giving students first-hand experience with scheduling policies, queue limits, and performance bottlenecks.

This paper details the structure of the assignment, presents an evaluation of three course offerings (Spring 2022, 2024, 2025), and discusses its pedagogical impact.

\section{Related Work}
Several works have also explored matrix multiplication as a vehicle for teaching parallel programming. Fietkiewicz demonstrated the educational value of recursive matrix multiplication in a parallel setting and suggested requiring students to perform more detailed efficiency analyses as part of the learning process~\cite{fietkiewicz2019student}. More recently, Bober and Bylina investigated teaching parallel programming on students' personal computers using matrix multiplication with MKL, OpenMP, and SYCL libraries, emphasizing performance comparisons across multiple frameworks~\cite{csedu25}.

Our work extends these efforts in two key ways. First, rather than focusing primarily on framework-level differences, we situate matrix multiplication in a production-grade HPC workflow using the HiPerGator supercomputer. This exposes students to job scheduling, queue management, and resource allocation, experiences that are rarely available in undergraduate curricula. Second, by requiring students to benchmark implementations in both Python and C across multiple threading paradigms, we highlight the contrast between language-level concurrency models and performance bottlenecks. The novelty of our approach lies in combining algorithmic simplicity with the central learning goal of HPC workflow proficiency, thereby deepening student understanding of Parallel and Distributed Computing (PDC) and bridging the gap between this conceptually challenging area of the undergraduate curriculum and real-world applications.

\section{Assignment Design}

During the assignment design, we intentionally chose matrix multiplication because it is a well-known common algorithm that is computationally intensive ~\cite{stothers2010complexity}. Its algorithmic simplicity is a key advantage, as it allows students to focus on the primary learning objectives: navigating a real High-Performance Computing (HPC) environment to gain a better understanding of parallel programming and multithreading.

The assignment does not focus on the implementation of the algorithm itself. The main goal is for students to understand the impact of multithreading on algorithm performance. This goal is achieved through exposing students to the process of designing experiments, managing resources on a real-world supercomputer via a job scheduler, and analyzing performance variations across different parallelization strategies.

This approach situates theoretical knowledge within a practical, project-based framework, moving students from abstract concepts to tangible skills grounded in the principles of discovery learning~\cite{adams2021teaching}.

\subsection{Learning Objectives}

Upon completing this assignment, students are expected to be able to:
\begin{itemize}
    \item \textbf{Implement and Compare} different parallel programming methods (POSIX threads, OpenMP) as well as understand their performance implications.
    \item \textbf{Manage HPC Workflows} by writing SLURM batch scripts, submitting jobs to a scheduler, and monitoring their execution on a remote HiperGator cluster.
    \item \textbf{Analyze Performance Trade-offs} by designing an experiment to measure how execution time is affected by matrix dimensions, programming language choice, thread count, and core allocation.
\end{itemize}

\subsection{Structure}

We structured the assignment as an open-ended investigation. Rather than providing a rigid set of instructions, we task students with an open-ended objective to investigate and report on the factors affecting the performance of matrix multiplication. This structure encourages a discovery-based approach where students must formulate their own hypotheses and design experiments to test them.

Students completed the project individually, ensuring that each student gained hands-on experience with both implementation and HPC workflows. However, we encouraged informal peer discussion during lectures, particularly when troubleshooting job scripts or interpreting unexpected performance results. 

To help students remain focused on the long-duration aspects of the project, we divided the work in the assignment instructions into discrete milestones. Each stage (implementation, HPC workflow, data analysis and final report) functioned as a mini-project with its own deliverables.

We structured the process as follows:
\begin{enumerate}
    \item \textbf{Baseline Implementation:} Students first implement standard, single-threaded matrix multiplication in Python and C. This establishes a performance baseline.
    \item \textbf{Parallel Implementation:} Students then develop two parallel versions in C (using POSIX threads and OpenMP) and one in Python (using its threading library).
    \item \textbf{HPC Immersion:} Students are given access to HiPerGator and its documentation. They learn to write shell scripts to automate their experiments and submit jobs to the SLURM scheduler. This is a critical task that mirrors the typical workflow of researchers and software engineers. They must specify job parameters such as core count, memory allocation, directly engaging with concepts of resource management discussed in lectures.
    \item \textbf{Data Collection and Analysis:} Students run their code with varying matrix sizes, thread counts, and core allocations. They record results and are required to produce visualizations to support their analysis.
    \item \textbf{Final Report:} The project culminates in a technical report where students present their findings, interpret their graphs, and explain the observed performance behaviors.
\end{enumerate}

\section{Evaluation and Discussion}

We evaluated the assignment's effectiveness using two primary methods: direct analysis of student work products (final reports and code) and indirect assessment via anonymous surveys measuring student perceptions. As shown in Table ~\ref{table:survey_summary}, data is aggregated from three semesters: Spring 2022, 2024, and 2025.

\subsection{Analysis of Student Submissions}
Students submitted final reports that provide clear evidence of student learning. By tasking students to explain their results, we can assess their conceptual understanding. The figures below, taken from a representative student report, are presented here with the kind of analysis we expect students to produce, which demonstrates the learning that occurred.

\begin{figure}[h!]
    \centering
    \includegraphics[width=0.9\textwidth]{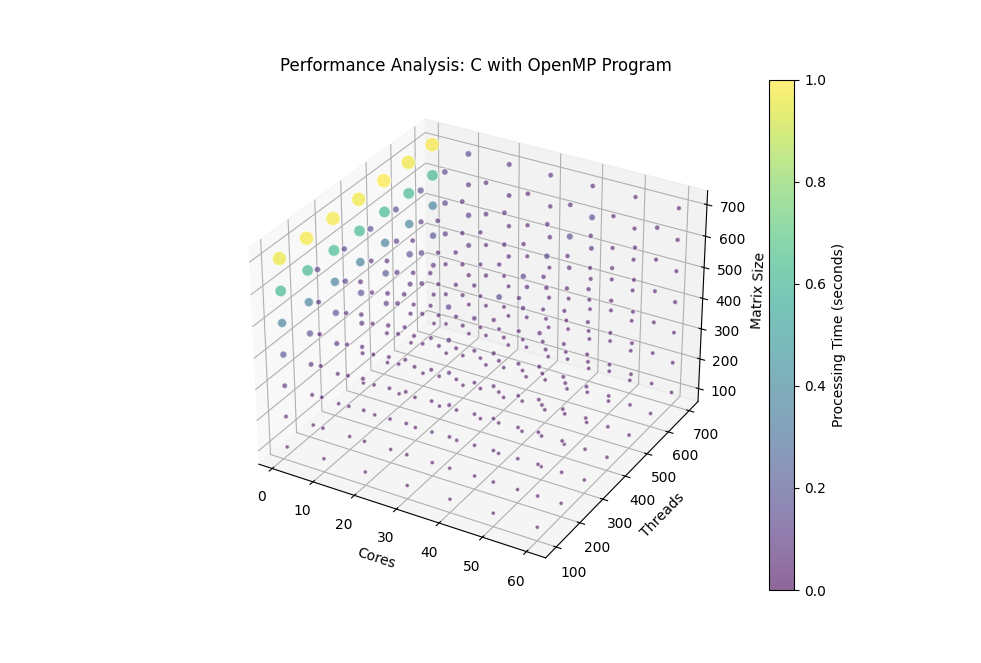}
    \caption{Performance of C with OpenMP.}
    \label{fig:omp}
\end{figure}

\begin{figure}[h!]
    \centering
    \includegraphics[width=0.9\textwidth]{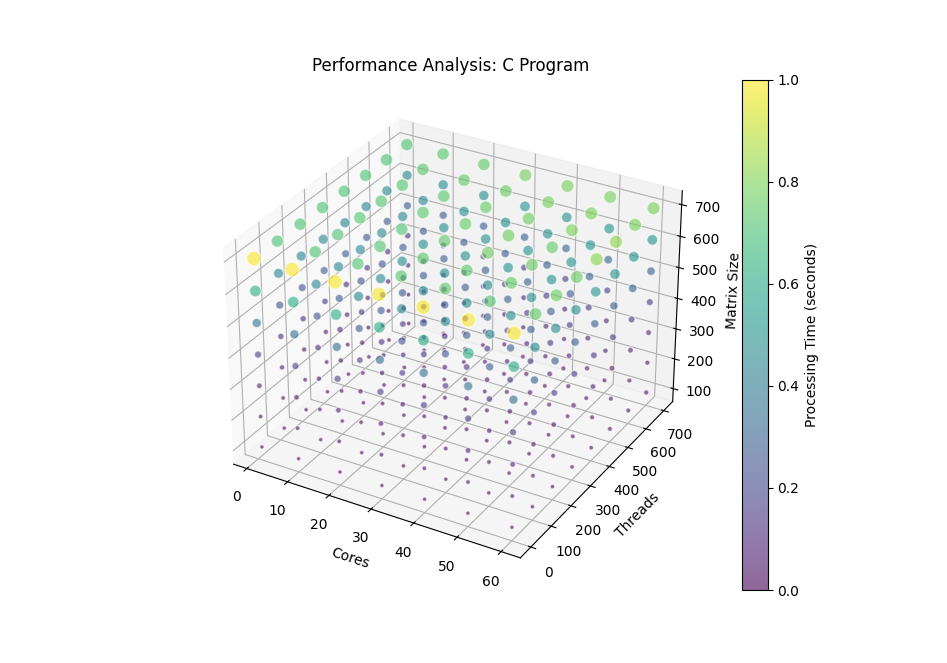}
    \caption{Performance of C with POSIX threads.}
    \label{fig:pthreads}
\end{figure}

Figures \ref{fig:omp} and \ref{fig:pthreads} show a student's comparison of C-based implementations. In their reports, students identified the C implementations as significantly faster than Python — as expected. They noted that the OpenMP version (Figure \ref{fig:omp}) offered the best performance with the least programming effort, providing a concrete example of the value of libraries that are optimized for parallel execution.

The slight performance differences and diminishing returns at higher thread counts led students to independently research and discuss concepts like thread creation overhead and the limits of parallelization.

\begin{figure}[h!]
    \centering
    \includegraphics[width=0.9\textwidth]{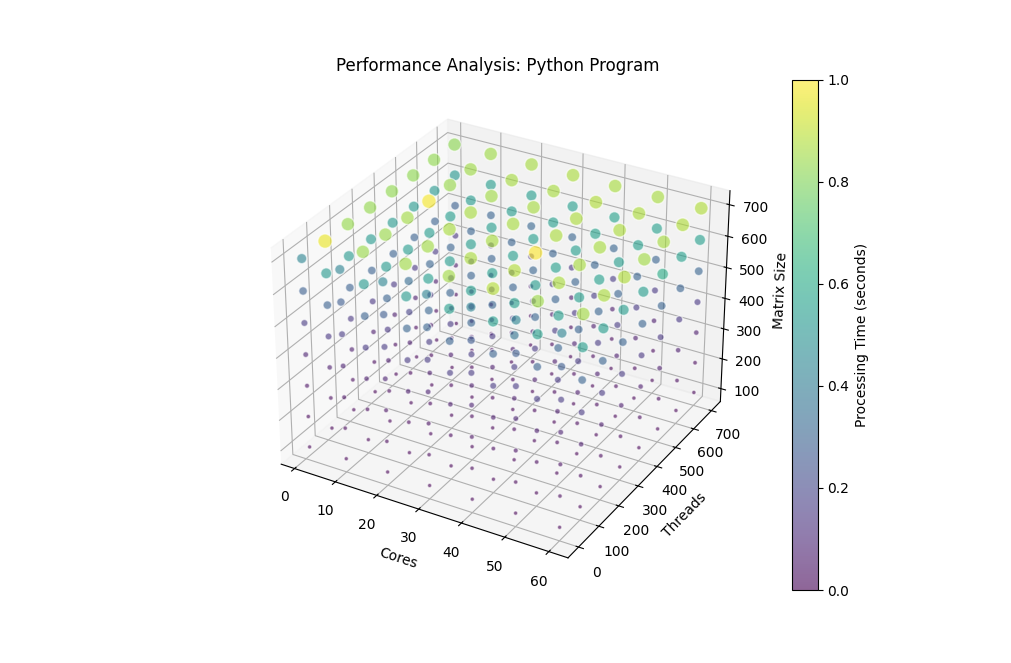}
    \caption{Performance of multithreaded Python.}
    \label{fig:py_threaded}
\end{figure}

A key insight for many students was the unexpected performance behavior of multithreaded Python. As shown in Figure \ref{fig:py_threaded}, increasing the number of threads did not lead to performance gains; in fact, some students observed slight slowdowns. This prompted students to re-examine their initial intuition that "more threads provide more speed", fostering a deeper understanding of the nuances of concurrency.

\subsection{Student Perceptions and Self-Reported Learning}

We collected Anonymous survey data at the end of each semester. Survey data analysis confirms that students found the assignment engaging, useful, and effective. As shown in Table \ref{table:survey_summary}, student reception has been overwhelmingly positive across all three course offerings.

\begin{table}[htbp]
\centering
\caption{Survey Results by Year\protect\footnotemark}
\label{table:survey_summary}
\resizebox{\textwidth}{!}{%
\begin{tabular}{lccccc}
\toprule
\textbf{Year} & \textbf{Interest} & \textbf{Understanding} & \textbf{Useful} & \textbf{Recommend} & \textbf{Difficulty} \\
 & \textbf{Y/M/N (\%)} & \textbf{Y/M/N (\%)} & \textbf{Y/M/N (\%)} & \textbf{Y/M/N (\%)} & \textbf{(1--5)} \\
\midrule
2022 & 91/9/0 & 73/27/0 & 100/0/0 & 91/9/0 & 3.27 \\
2024 & 75/25/0 & 88/0/12 & 63/37/0 & 88/12/0 & 3.50 \\
2025 & 64/29/7 & 71/22/7 & 79/14/7 & 79/14/7 & 3.79 \\
\bottomrule
\end{tabular}%
}
\end{table}
\footnotetext{The course was not offered in Spring 2023 due to instructor sabbatical. Y/M/N represents the percentage of Yes/Maybe/No responses for each survey question.}

The consistently high ratings for usefulness and willingness to recommend the project indicate that students recognize its value. 

Qualitative feedback reveals three consistent themes across all cohorts. When asked \textbf{``What did you like most about the project?''}, students responded:

\textbf{Access to Professional Computing Resources:}
Students consistently valued the opportunity to work with production HPC infrastructure:
\begin{itemize}
    \item ``Being able to use a supercomputer to run extremely large computations with a large amount of memory'' (2022)
    \item ``Using a super computer is something that I can put on my resume'' (2022) 
    \item ``Getting to use UF's HiPerGator was an experience that not a lot of undergraduates get to use'' (2025)
\end{itemize}

\textbf{Hands-on Performance Observation:}
Students appreciated seeing theoretical concepts manifest in measurable performance differences:
\begin{itemize}
    \item ``I loved seeing the results across languages. It's fascinating to see just how slow python is for stuff like this'' (2024)
    \item ``I liked that the program made it very clear how threading effects time needed for execution of programs'' (2022)
    \item ``I enjoyed being able to see the difference in my home laptop vs what a supercomputer is capable of'' (2025)
\end{itemize}

\textbf{Applied Learning Experience:}
Students valued the practical application of classroom theory:
\begin{itemize}
    \item ``I liked the hands on approach that the project afforded me'' (2022)
    \item ``Actually writing code to see how the discussed topics work in real life scenarios'' (2024)
\end{itemize}

This feedback, combined with the quantitative data, supports the effectiveness of providing students with hands-on HPC experiences for learning PDC concepts.

\section{Conclusion}

This paper presents a project-based learning assignment that leverages a production grade supercomputer to teach fundamental concepts in parallel and distributed computing (PDC). By framing a classic matrix multiplication problem within a discovery-learning framework, we successfully moved students beyond rote memorization of algorithms to a practical understanding of real-world HPC workflows, performance bottlenecks, and language-specific concurrency models.

The primary contribution of this work is a replicable pedagogical model that addresses the common challenge of teaching abstract PDC concepts. The analysis of student work demonstrates clear learning gains in understanding multithreaded behavior and HPC workflows. Furthermore, student self-reported data confirms the assignment is highly engaging and valuable. By situating theoretical knowledge within a practical, hands-on context, students not only learn the material more deeply but also acquire practical skills relevant to skills used in the industry. This approach can be replicated to provide an effective PDC education at the undergraduate level.
\medskip

\section{Acknowledgments}
We thank Jonathan Lamoureux and Skyler Gipson for their contributions to earlier stages of this paper. 

The authors also acknowledge UFIT Research Computing for providing computational resources and support that have contributed to the research results reported in this publication. URL: \url{http://www.rc.ufl.edu}

\printbibliography

\end{document}